\documentclass[
    ,final            
  ]
  {aipproc}

\layoutstyle{6x9}

\begin{document}

\title{MSSM Dark Matter Without Prejudice}
\rightline{\vbox{\halign{&#\hfil\cr
&SLAC-PUB-13799\cr
}}}

\classification{11.30Pb,12.60.Jv,14.80.Ly,95.35.+d}
\keywords      {pMSSM, general MSSM phenomenology, dark matter}

\author{James S. Gainer\footnotemark}{
  address={SLAC National Accelerator Laboratory, 2575 Sand Hill Rd., 
Menlo Park, CA, 94025, USA}
}

\begin{abstract}
Recently we examined a large number of points in a 19-dimensional
parameter subspace of the CP-conserving MSSM with Minimal Flavor
Violation.  We determined whether each of these points satisfied existing
theoretical, experimental, and observational constraints.  Here we discuss the
properties of the parameter space points allowed by existing data that
are relevant for dark matter searches.
\end{abstract}

\maketitle

\footnotetext{Present addresses: Department of Physics and Astronomy,
  Northwestern University, Evanston, IL 60208 \\
  High Energy Physics Division, Argonne National Laboratory, Argonne,
  IL 60439}


\section{Introduction}

The MSSM has a large parameter space; this raises the question of how
well we know its properties aside from specific SUSY breaking
scenarios such as mSUGRA, AMSB, GMSB, etc.
In an attempt to address this question, we performed 
scans of a 19-dimensional subspace of the full 100+ parameter
(R-parity conserving) MSSM (sometimes referred to as the
``phenomenological MSSM'') and applied a comprehensive set
of theoretical, experimental, and observational constraints, thereby
obtaining a large set of ``models'' (parameter space points) which are
consistent with existing data\cite{us}.  
The particulars of and results from this scan are discussed in other
presentations at this conference\cite{tom,john}.  Here we will be
concerned with the implications of this scan for dark matter,
addressing the properties of LSPs in the set of allowed MSSM models
generated, and in particular their signatures in direct detection and
indirect detection experiments. 

\section{Properties of the LSP}

\begin{figure}[htbp]
  \includegraphics[width=11.0cm,angle=0]{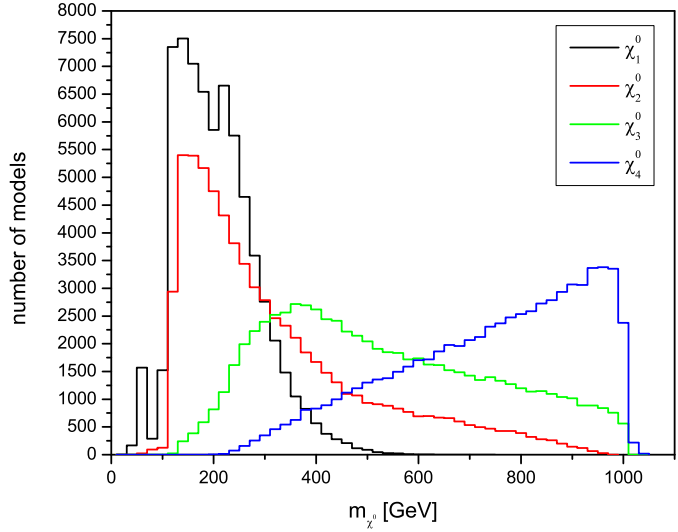}
  \caption{The distribution of neutralino masses for models in the 
    model set.}
  \label{fig1}
\end{figure}

A histogram of the masses of the four neutralino species in the model set
is shown in Figure~\ref{fig1}. In all models in this set, the lightest
neutralino is the LSP.  

It is also notable that most LSPs are relatively pure
eigenstates, with models where the LSP is Higgsino or mostly Higgsino
being by far the most common.  A precise description of the content of
LSPs in the model set is presented in Table~\ref{neutralino mixing}.
The prevalence of nearly pure eigenstates is not surprising;
one would expect the LSP be a pure eigenstate fairly
often in a random scan of Lagrangian parameters such as that
performed.  As the differences between $M_1,M_2,$ and $\mu$ will often
be large compared to $M_Z$, the eigenstates of the resulting mixing
matrix will be essentially pure gaugino and Higgsino states.

\begin{table}
\centering
\begin{tabular}{|l|c|r|} \hline\hline 
LSP Type & Definition & Fraction \\ 
& & of Models \\ \hline
Bino & $|Z_{11}|^2 > 0.95$ & 0.14 \\
Mostly Bino & $0.8 < |Z_{11}|^2 \leq 0.95$ & 0.03 \\
Wino & $|Z_{12}|^2 > 0.95$ & 0.14 \\
Mostly Wino & $0.8 < |Z_{12}|^2 \leq 0.95$ & 0.09\\
Higgsino & $|Z_{13}|^2+|Z_{14}|^2 > 0.95$ & 0.32 \\
Mostly Higgsino & $0.8 < |Z_{13}|^2+|Z_{14}|^2 \leq 0.95$ &  0.12 \\
All other models & & 0.15 \\
\hline\hline
\end{tabular}
\caption{The fractions of models in the model set for which the LSP is of each
  of the given types.  These types are defined in terms of the modulus squared
  of elements of neutralino mixing matrix in the SLHA convention.}
\label{neutralino mixing}
\end{table}

\section{Relic Density}

\begin{figure}[htbp]

\includegraphics[width=5.0cm,angle=-90]{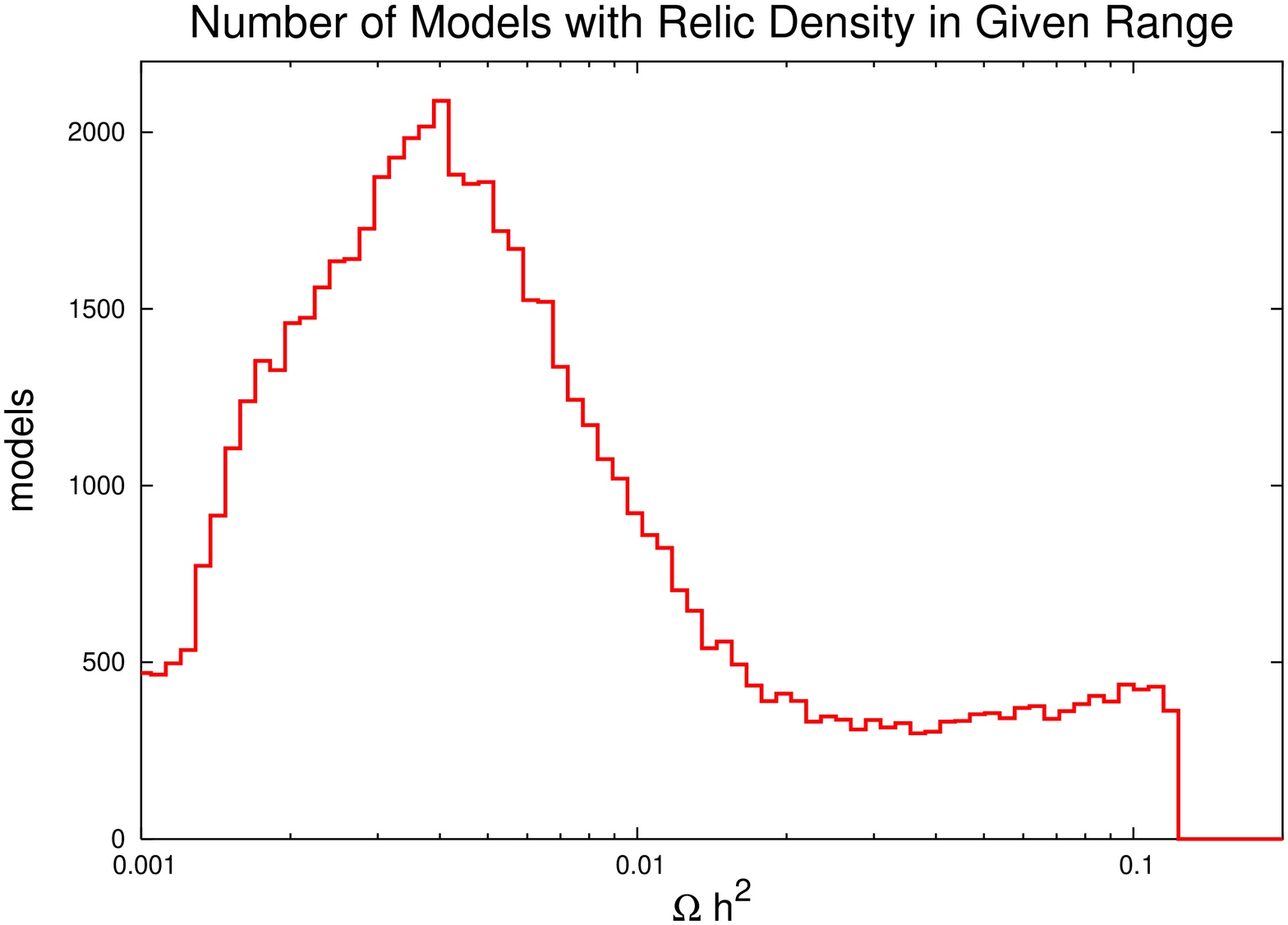}
\hspace*{-0.1 cm}
\includegraphics[width=5.0cm,angle=-90]{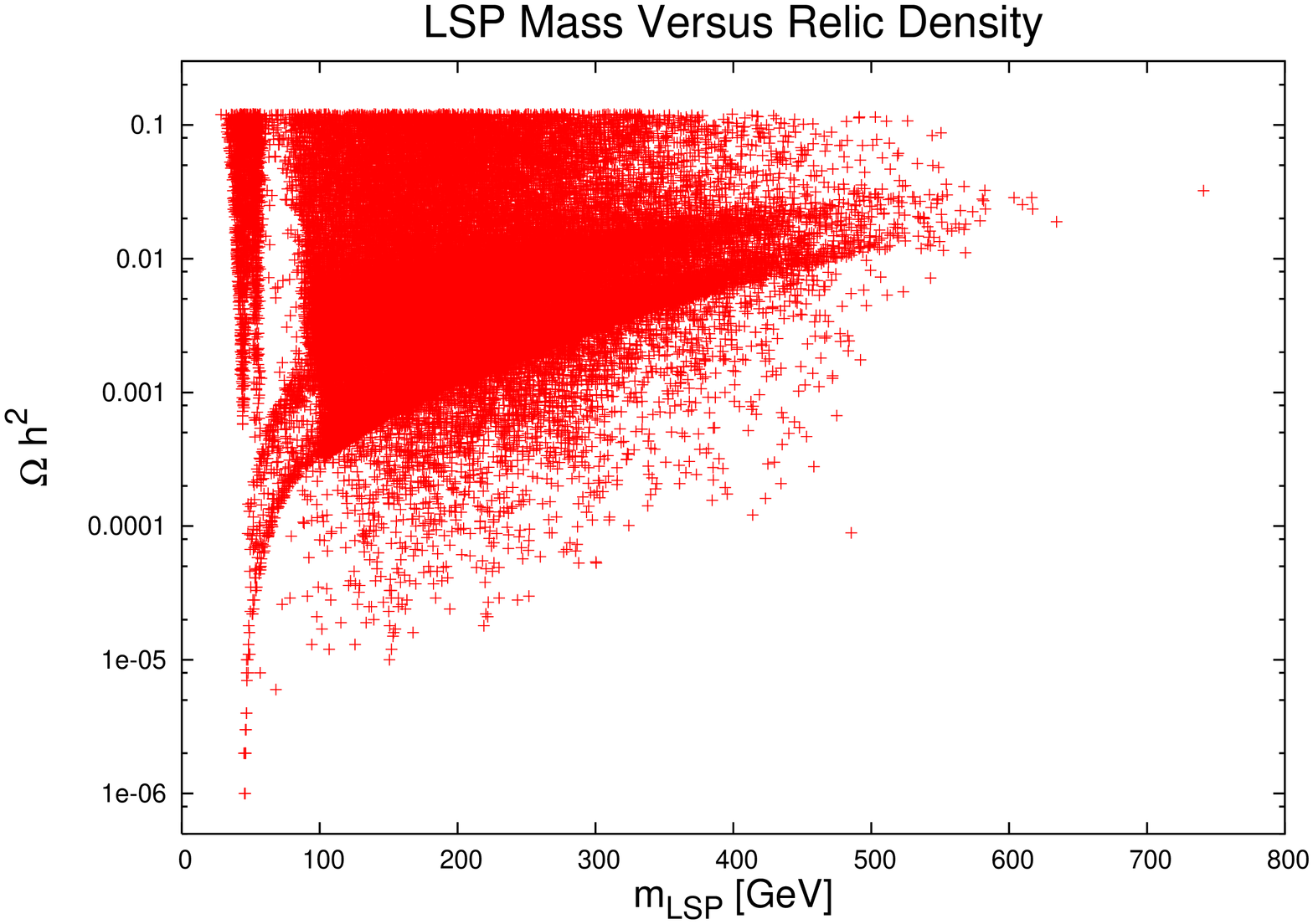}
\caption{
The left panel shows the distribution of $\Omega
h^2|_{\mathrm{LSP}}$ for models in the model set.  The right panel
shows the values of $\Omega h^2|_{\mathrm{LSP}}$ versus LSP mass for
models in the model set.}
\label{fig2}
\end{figure}

In applying constraints to models, we did not demand that the LSP, in
any given model, account for all of the dark matter; we required only
that the (thermal) LSP relic density not be too large to be consistent
with WMAP.
The distribution of $\Omega h^2|_{\mathrm{LSP}}$ values in our model
set is shown in the left panel of Figure~\ref{fig2}.  
It is interesting to note that this distribution is peaked at small
values of $\Omega h^2|_{\mathrm{LSP}}$ and that the range of possible
values of $\Omega h^2|_{\mathrm{LSP}}$ is found to be much larger than
those obtained by analyses of specific SUSY breaking scenarios.

The distribution of predictions for  $\Omega h^2|_{\mathrm{LSP}}$ as a
function of the LSP mass for models in the model set is shown in the
right panel of Figure~\ref{fig2}.
We see from this figure that $\Omega h^2|_{\mathrm{LSP}}$ generally increases 
with the LSP mass.  However, a large range of values for the relic
density are possible at any given LSP mass.  
The empty region in Figure~\ref{fig2} where $\Omega
h^2|_{\mathrm{LSP}} \approx 0.001 - 0.1$ and $m_{\mathrm{LSP}}\approx
50-100$ is due to the paucity of models with Higgsino or Wino LSPs in
this mass range (as such models would generally have a chargino light
enough for discovery at LEP) together with the fact that in general,
LSPs which are mostly Higgsino or Wino give lower values of $\Omega
h^2|_{\mathrm{LSP}}$.

\section{Direct Detection of Dark Matter}

To implement constraints on MSSM parameter space resulting from direct
WIMP detection experiments and to study the direct detection
signature of allowed models, we calculated the spin-dependent and
spin-independent WIMP-nucleon cross sections using micrOMEGAs 2.21.
The quantity actually measured in experiments is the WIMP-nucleon cross
section scaled to the fraction of the dark matter density represented
by the LSP, hence the cross section data presented in
the figure below are scaled by $\xi = \Omega
h^2|_{\mathrm{LSP}} / \Omega h^2|_{\mathrm{WMAP}}$.  
These experiments generally provide a more significant bound on the
spin-independent WIMP-nucleon cross sections, and hence we will focus
on these.

\begin{figure}[htbp]

\includegraphics[width=5.0cm,angle=-90]{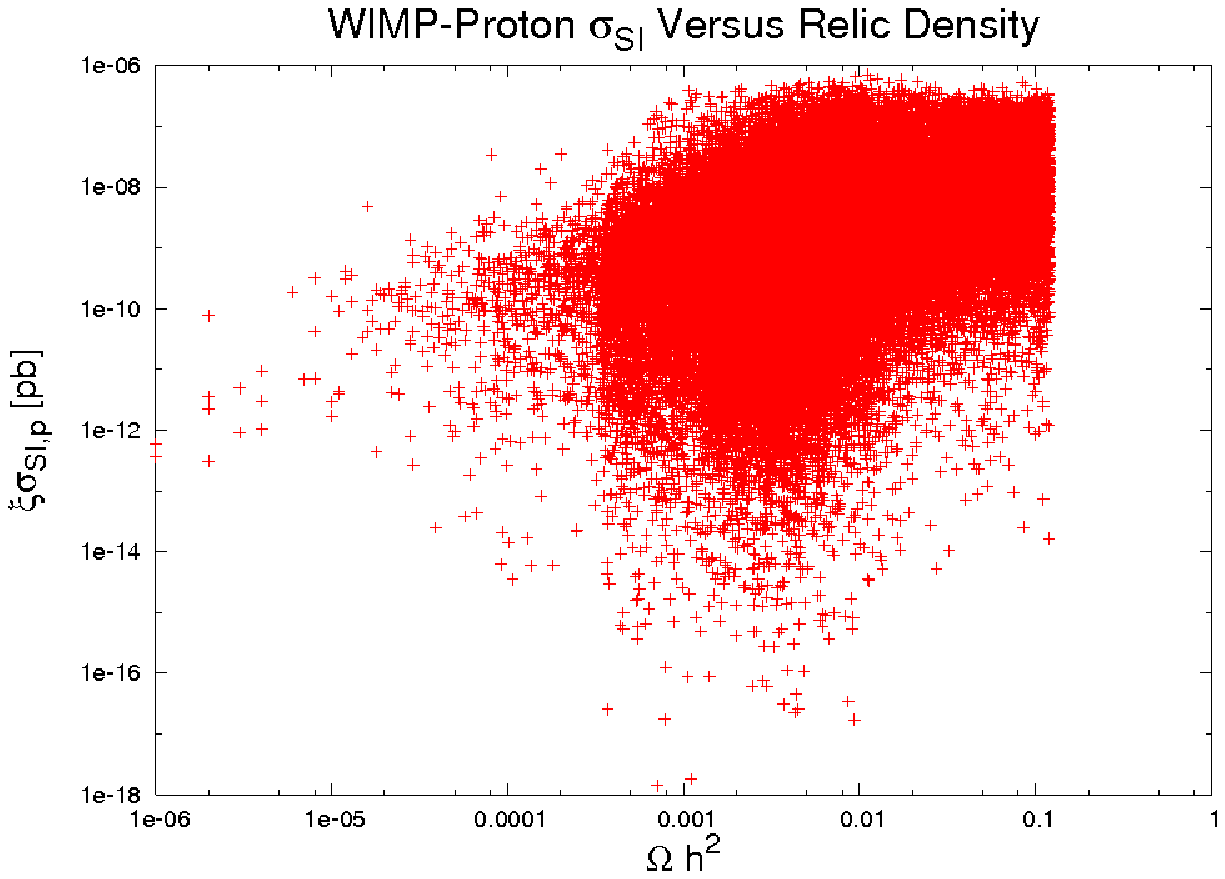}
\hspace*{-0.1 cm}
\includegraphics[width=5.0cm,angle=-90]{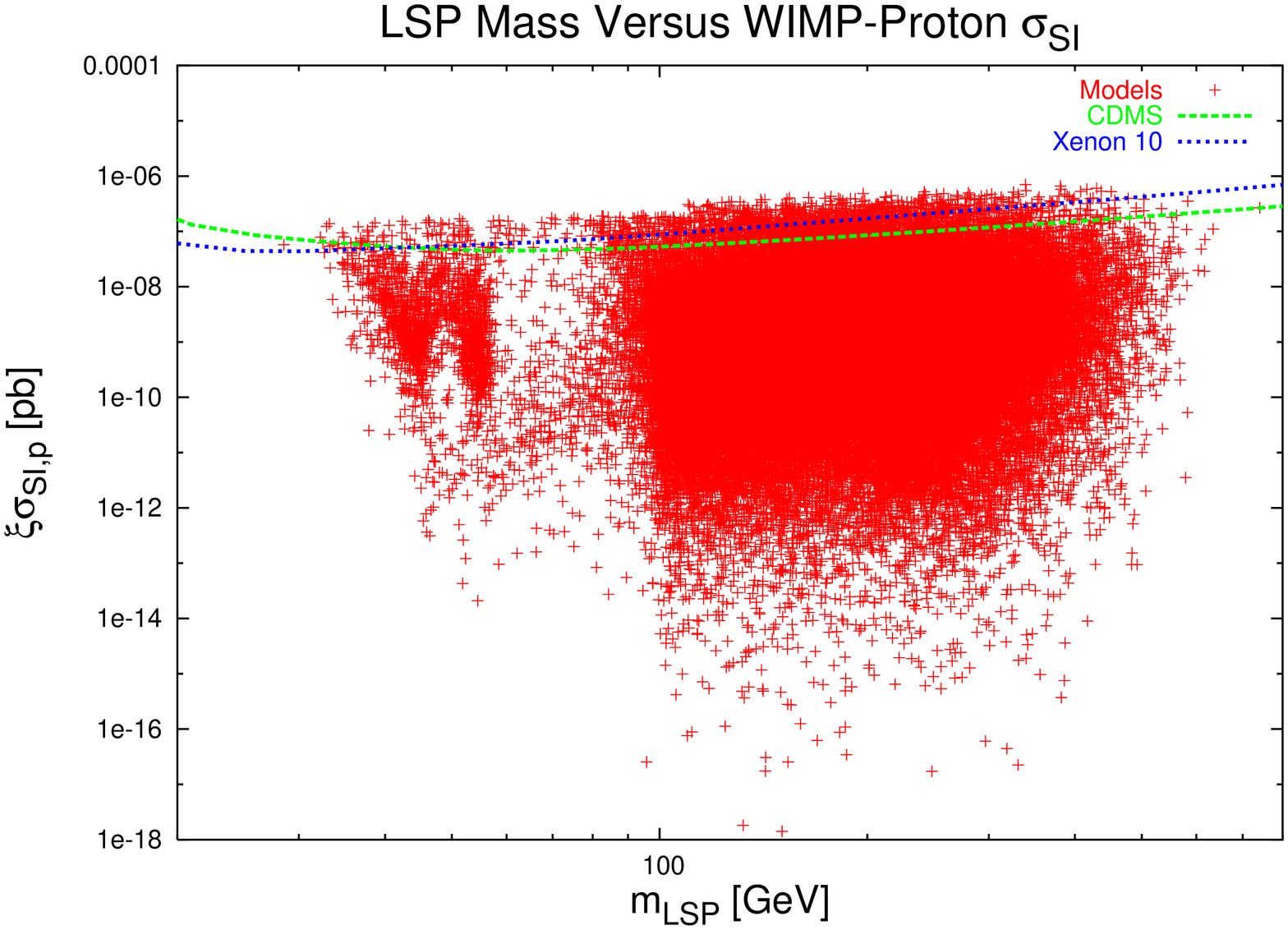}
\caption{
The left panel shows the distribution of scaled WIMP-proton spin-independent cross section
versus the LSP contribution to relic density for models in the model
set, while the right hand panel shows the values of scaled WIMP-proton
spin-dependent cross versus LSP mass for models in the model set.  In the
right panel, the CDMS and Xenon10 bounds, which provide the strongest
limits for the range in LSP mass relevant for these models, are shown.}
\label{fig3}
\end{figure}

In the left panel of Figure~\ref{fig3}, the distribution for the scaled WIMP-proton
spin-independent cross section versus relic density for our model
sample is presented.  Perhaps not surprisingly, larger values of the
cross section are generally found at larger values of $\Omega
h^2|_{\mathrm{LSP}}$.  
However, we note that even for relic densities close to the
  WMAP value, where there is little contribution to the diversity in
  scaled cross section from variation in the relic density, $\xi
  \sigma_{p,SI}$ is seen to vary by almost eight orders of magnitude.

In the right panel of Figure~\ref{fig3}, we see the scaled
WIMP-proton spin-dependent and spin-independent cross sections as a
function of the LSP mass, together with the constraints from XENON10 and CDMS.
To take into account significant theoretical uncertainties in the
calculation of the WIMP-proton cross section, we allowed for a factor
of 4 uncertainty in the calculation of the WIMP-nucleon cross
section.  Here as well, the range in the value of the scaled
WIMP-nucleon cross section is notable.

\subsection{Indirect Detection of Dark Matter}

The PAMELA collaboration has recently claimed an excess in the ratio
of cosmic ray positrons to electrons observed at energies above
$10$ GeV.  An attempt to quantify the extent to which these results,
together with various other observations including those of the
Fermi-LAT, may be understood in terms of LSP annihilation in SUSY
models in the model set described here is ongoing; some early results
have been presented\cite{njp}.


\begin{theacknowledgments}
Work supported in part by the Department of Energy, Contract DE-AC02-76SF00515.
\end{theacknowledgments}


\begin{thebibliography}{9}
\bibitem{tom}
The procedure by which this model set was generated is also described in
more detail in the presentation by T.~G.~Rizzo:
 arXiv:0907.1668 [hep-ph].
\end{thebibliography}
\end{document}